\documentclass[twocolumn,showpacs,preprintnumbers,amsmath,amssymb]{revtex4}

\usepackage{graphicx}
\usepackage{dcolumn}
\usepackage{bm}
\usepackage{amssymb}

\newcommand\nuh{\nu_h \rightarrow \gamma  \nu}

\newcommand\pair{e^+ e^-}
\newcommand\mix{|U_{\mu h}|^2}

\def\address{\@ifstar{\address@star}%
  {\@ifnextchar[{\address@optarg}{\address@noptarg}}}

\begin{document}

\author{S.N.~Gninenko}

\affiliation{Institute for Nuclear Research, Moscow 117312}


\title{Comments on arXiv:1011.3046 "Muon Capture Constraints on Sterile Neutrino Properties"}
\date{\today}
\begin{abstract}
It has been very recently reported (McKeen, Pospelov, arXive:1011.3046) that the parameter region suggested for  an
 explanation of the  neutrino oscillation results  from the LSND, KARMEN and MiniBooNE experiments in terms of the production and   radiative decay of a heavy neutrino ($\nu_h$) can be ruled out based on the measurements of the radiative muon capture (RMC) on hydrogen. We calculate limits on mixing strength between the $\nu_h$ and $\nu_\mu$ by using results of this experiment, and  find out that they essentially disagree  with the reported bounds. For the $\nu_h$ with mass 60 MeV  our limit is worse by a factor of 5,  
while for 90 MeV it is worse by about two orders of magnitude. We also noticed the wrong behavior of the reported limit curve in the low mass region. The importance of accurate Monte Carlo simulations of the $\nu_h$ signal in the RMC experiment is stressed.
Our conclusion is that the whole LSND-MiniBooNE parameter region cannot be  ruled out by the RMC measurements.
\end{abstract}
\maketitle
In the recent work \cite{sng1}, see also \cite{sng2}, it has been shown  that the neutrino oscillation results 
 from the LSND, KARMEN and MiniBooNE experiments   
could be explained by  the existence of a  heavy sterile neutrino ($\nu_h$),  assuming  
 that it is  created by mixing in  $\nu_\mu$ neutral-current interactions and  decays radiatively  into  
a photon and a light neutrino. The  $\nu_h$'s could be Dirac or Majorana type and  decay  
dominantly into $\nuh$ in the detector target if, for example,  there is a non-zero transition
magnetic moment between the $\nu_h$ and the
$\nu$. Combined analysis of the 
energy and angular distributions of the LSND and MiniBooNe excess events 
suggests that the $\nu_h$  mass $m_{\nu_h}$, the mixing strength  $|U_{\mu h}|^2$
and the lifetime  $\tau_{\nu_h}$, are in the ranges:
\begin{eqnarray}
 40 \lesssim m_{\nu_h} \lesssim 80~ {\rm MeV},~ 10^{-3}\lesssim |U_{\mu h}|^2 \lesssim 10^{-2}, \nonumber\\
10^{-11}\lesssim  \tau_{\nu_h} \lesssim 10^{-9}~s,
\label{param} 
\end{eqnarray}
 respectively.

The  mixing  $|U_{\mu h}|^2$ for the mass range of \eqref{param}  
would result in the $\nu_h$ emission in the ordinary muon capture (OMC) 
on nuclei $\mu^- A \to \nu_\mu A'$ \cite{deu}.
The OMC rate of the heavy neutrino production can be estimated as  
\begin{equation}
 \Gamma_{\nu_h} = \Gamma_{OMC} \mix \rho(m_{\nu_h})/\rho(0) 
\end{equation}
where $\Gamma_{OMC} $ is the OMC rate 
and $\rho(m_{\nu_h}),~\rho(0)$ are the phase space factors
for emission of heavy neutrino and  massless, unmixed neutrinos, respectively.\\  
Recently, it has been  noticed by McKeen and Pospelov \cite{pospel} 
that the parameter space of \eqref{param} can be probed by using 
the results of the experiment  on the radiative muon capture (RMC) rate on hydrogen \cite{rmc}. 
In this experiment muons were stopped in a liquid hydrogen target. Photons from the reaction 
 $\mu p \to \nu_\mu \gamma n$ were 
 converted in a Pb layer surrounding the target into  $\pair$ pairs , whose momenta 
 were measured by a magnetic spectrometer. If  the $\nu_h$ exists, it would be produced through mixing in the OMC process and after  
the prompt $\nuh$ decay in the target and subsequent decay photon conversion result in a final state identical to the one from 
the RMC reaction. 
\begin{figure}[tbh!]
\begin{center}
    \resizebox{8cm}{!}{\includegraphics{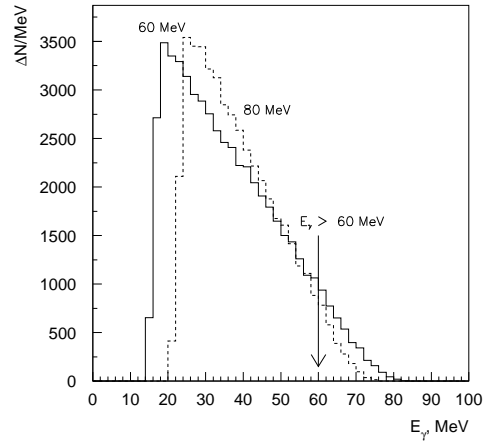}}
     \caption{ Energy distributions of photons from the  $\nuh$ decay of the Dirac heavy heutrino with $a=-1$ \cite{sng1} with energy of 100 MeV
     created through the mixing in the muon capture $\mu_\nu p \to \nu_h n$ on proton and  
      calculated for  $\nu_h$ masses of 60 and 80 MeV. The arrow indicate  the threshold of 60 MeV
      used in the analysis of RMC data \cite{pospel}. }
\label{spectrum}
\end{center}
\end{figure}
The number of RMC photons observed in the experiment is calculated 
by using Eq.(38) from  \cite{rmc}:
\begin{equation}
 n_{\gamma} = n_{stop} R_{\gamma} K A_a 
 \label{rmc}
\end{equation}
where  $n_{stop} (\simeq 3.4 \cdot 10^{12})$ is the total number of  muons stopped in the target,
$R_\gamma = (2.1\pm 0.21)\cdot 10^{-8}$ is the RMC branching fraction, K = 0.6 is an effciency factor, and 
 $A_a$ is the absolute photon spectrometer acceptance  which include 
geometrical acceptance, conversion probability, $\pair$ reconstruction efficiency etc..
Using  $n_{\gamma} = 279 \pm 26$ events found in the experiment, gives averaged over the RMS photon spectrum value $A_a = 6.5 \cdot 10^{-3}$.
We will use these numbers for the cross-check, but immediately note that the $A_a$ value  
is about two times bigger than that estimated for the $\nuh$ spectrum. To be conservative we will keep this "safety" factor just in mind.\\
The energy distribution of photons 
from the $\nuh$ decay calculated for different masses of the $\nu_h$ 
with the total energy of 100 MeV, which is expected from the OMC on hydrogen, is shown in Fig.\ref{spectrum}
The  upper limit on the  excess number of photons from the 
$\nuh$ decay in the RMC experiment can be estimated at 2$\sigma$ level 
for  $E_\gamma> 60 $ MeV as:
\begin{equation}
 \Delta n_{\gamma} = n_{\gamma}-n_{0} \lesssim 140~š{\rm events}  
\label{events}
\end{equation}
where  $n_{0} \simeq 180$ is the number of events that are expected 
from the theory for the ratio of the pseudoscalar and axial-vector form factors $g_p/g_a=0.69 $ and ortho to para transition rate 
of the ($\mu p$) atom equal $\lambda= 4.1\times 10^4$ s$^{-1}$ \cite{rmc}.\\  
The excess $n_s$  of events from the $\nuh$ decays can be calculated as 
\begin{equation}
 n_{s} = n_{stop} R_{OMC} \mix f_{phs}  B(\nuh) P_{\nuh} f_\gamma K A_a
\label{excess}
\end{equation}
where $n_{stop},~K,~ A_a$ are as in \eqref{rmc}, $R_{OMC} (1.5\cdot 10^{-3})$ is 
the OMC branchning fraction, $B(\nuh)(\simeq 1)$ is the branching fraction of the $\nuh$ decay, 
$f_{phs} = \rho(m_{\nu_h})/\rho(0)$ is the ratio of phase space factors,
$P_{\nuh} (0.15)$ is the probability to decay in the  target for heavy neutrino lifetime $\tau = 10^{-9}$ s
as in \cite{pospel},
 $f_\gamma$ is the fraction of photons 
with energy $E_{\gamma} > 60$ MeV (0.074 and 0.05 for 60 and 80 MeV $\nu_h$, respectively)  . 
Taking into account  above values results in the 2$\sigma$ limit  $\mix < 1.4\times 10^{-3}$ 
for the $\nu_h$ mass of 60 MeV, which is worse than the limit of \cite{pospel} by a factor of 5, see Fig. 3 in \cite{pospel}.
For the mass of 80 MeV the limit is $\mix < 4\times 10^{-3}$ and the disagreement is about factor 10, 
but for 90 MeV it is two orders of magnitude,
 $\mix < 5.1\times 10^{-2}$ vs  $\mix \lesssim 5\times 10^{-4}$ reported in \cite{pospel}.
 For the mass range $m_{\nu_h} \lesssim 20$ MeV the behavior of the limit curve shown in Fig.3 \cite{pospel} and its 
 extrapolation  to zero mass  is incorrect , because the probability $P_{\nuh} \to 0$ with  $m_{\nu_h}\to 0$.\\    
Let us note that work of Ref.\cite{sng1} has  been  attempted to explain an excess of events observed by   
the LSND in the  energy range $20 < E_\gamma < 60 $ MeV. The idea is to introduce a new particle with such 
decay properties that would allow to keep the number of events with the energy deposition above 60 MeV close to zero, in agreement with the  LSND observations.
This is achieved by suggesting that the $\nu_h$ decays radiatively as a Dirac particle with the photon asymmetry
parameter equal to $a= -1$. The emission of photons preferably  backward with respect to the $\nu_h$ direction of move 
makes their energy spectrum much softer compare e.g. to the one form the isotropic distribution 
(probably used in the analysis of Ref.\cite{pospel}).
In contrary the limits of Ref.\cite{pospel} are extracted from the data of the RMC experiment for the energy region 
$E_\gamma > 60$ MeV and even for the lower $\nu_h$ energy than in the LSND case. The fraction of expected $\nuh$ events in this energy range is small, and hence the limits obtained are quite sensitive to the details of the experimental analysis, in particular for the $\nu_h$ masses above 60 MeV. 
 Let us give an example of such sensitivity related to the photon response function in the RMC experiment.
 It is determined by generating Monte Carlo photons sampled from the known $\pi^- p$ reaction
spectrum, whose origins in the target are identical with the pion stopping distribution. The shape of this 
function is important to extract correct number of RMC events. The function is  essentially asymmetric ( see Fig. 10), and shifts
the whole energy spectrum  to the low energy region. The peak value of 129 MeV is shifted by 5-7 MeV but the low energy tail has
energies up to 50\%  of the initial photon energy. If we shift  spectra shown in Fig.1 just by, say   4 MeV,  the 
 decrease in the number of observed $\nuh$ events would result in limits $\mix <2.1\cdot 10^{-3}$ and $\mix < 0.01$ for the 60 and 80 MeV $\nu_h$, respectively, instead of given above. Therefore, 
accurate Monte Carlo simulations of the $\nu_h$ production 
and decay sequence events in the experiment \cite{rmc} and propagation of theses events through 
the spectrometer taking into account the response function is important to extract 
reliable limits on $\mix$. \\ 
The Primakoff mechanism of the $\nu_h$ production $\nu  A \to \nu_h A$ followed by the decay $\nu_h \to \gamma \nu$
 through the transition magnetic moment $\mu_{tr}$ between the $\nu_h$ and light $\nu$
with the subsequent $\nuh$ decay has been originally considered in \cite{gk} and has been used to 
extract limits on $\mu_{tr}$ from the NOMAD data \cite{nomad}. 
In order to use this mechanism for interpretation of the
LSND $\nu_\mu$ data \cite{pospel}, one has to make an additional assumption that the  
 light $\nu$ is  a component of the $\nu_\mu$. Although this is possible, 
it requires introduction of additional unknown parameters.

  In summary, we do not find arguments of McKeen and Pospelov  \cite{pospel} convincing enough to exclude   
the whole LSND-MiniBooNE parameter region, at least for  Dirac scenario of \cite{sng1}.
 In contrary, their work enhances  motivation for a
more sensitive direct search for the $\nuh$ decay in a dedicated experiment.\\
I would like to thank D.S. Gorbunov for discussion and pointing out Ref. \cite{pospel} to me.

\end{document}